# Influence of continuous predictor modelling methods on prediction stability in clinical prediction model development: an empirical comparison using real clinical data

Phichayut Phinyo[1], Pakpoom Wongyikul[1], Noraworn jirattikanwong[1], Natthanaphop Isaradech[2], Wuttipat Kiratipaisarl[2], Suppachai Lawanaskol[3], Noppadon Seesuwan[4], Wachiranun Sirikul[2]

[1] Department of Biomedical Informatics and Clinical Epidemiology (BioCE), Faculty of Medicine, Chiang Mai University, Chiang Mai, Thailand.

[2] Department of Community Medicine, Faculty of Medicine, Chiang Mai University, Chiang Mai, Thailand.

[3] Chaiprakarn Hospital, Chiang Mai, Thailand.

[4] Department of Emergency Medicine, Lampang Hospital, Lampang, Thailand.

**Correspondence:** Wachiranun Sirikul; Department of Community Medicine, Faculty of Medicine, Chiang Mai University, Chiang Mai, Thailand. Email: wachiranun.sir@cmu.ac.th.

## Abstract

**Background and objective:** Prediction stability is increasingly recognised as important for reliable clinical prediction model development, but the effect of continuous predictor modelling choices remains unclear. This study examined how different approaches to modelling continuous predictors influence prediction stability.

**Methods:** We used a real clinical dataset of 19,418 emergency department patients, previously used to develop a prediction model, to create five sample size scenarios ranging from 437 to 8,739 patients. Six continuous predictor modelling methods were compared: dichotomisation at the median (DIC), tertile categorisation (TER), linear terms (LIN), quadratic terms (QUA), multivariable fractional polynomials (MFP), and extreme gradient boosting (XGB). Prediction stability was evaluated using a bootstrap-based framework. Optimism-corrected AUC and calibration measures were estimated through internal validation. A modelling approach was considered stable when at least 90% of individual predictions had a mean absolute prediction error (MAPE) ≤5%.

**Results:** Prediction stability increased with sample size but differed by modelling method. At n = 437, no method achieved the stability criterion; LIN was the most stable, followed by DIC. At the base sample size (n = 874), DIC and LIN achieved stable predictions with similar calibration, although DIC had lower AUC. At n = 1,748, QUA achieved stable predictions, while MFP and XGB did not. At larger sample sizes (n = 3,496 and n = 8,739), all methods achieved stability. LIN, QUA, MFP, and XGB generally had higher AUCs than DIC and TER, while XGB showed the highest AUC but persistent miscalibration.

**Conclusion:** In this empirical investigation, continuous predictor modelling methods appeared to influence prediction stability. LIN achieved stable predictions from the base sample size onwards, whereas QUA, MFP, and XGB required larger samples. Although XGB showed high discrimination, calibration concerns persisted. These findings suggest that, in smaller datasets, simpler approaches, particularly LIN, may provide more stable predictions.

**Keywords:** Prognosis; Risk prediction; Sample size; Prediction model; Accuracy

## Introduction

Clinical prediction models (CPMs) estimate the probability of health outcomes for individual patients using statistical or machine learning methods [1]. They are intended to support clinical decision-making by informing diagnosis, prognosis, treatment selection, and care planning. By providing individualised risk estimates, CPMs have considerable potential to support precision medicine through patient stratification and tailored clinical management [1, 2]. However, their uptake in routine practice remains limited. This implementation gap is partly attributable to concerns about methodological quality, including inadequate sample sizes, suboptimal predictor handling, insufficient validation, and incomplete reporting [3–5]. Consequently, questions remain about whether many published CPMs are sufficiently reliable and reproducible for clinical use.

Traditionally, CPM evaluation has focused on measures of overall predictive performance, particularly discrimination and calibration [6]. Although these measures remain essential, they do not fully describe whether a modelling approach produces reproducible individual-level predictions during model development [7, 8]. Prediction instability occurs when models developed using the same intended approach generate meaningfully different predictions across samples drawn from the same target population. This issue is particularly important for clinical decision-making because unstable individual risk estimates could lead to different decisions for similar patients [7, 9]. Prediction stability has therefore been increasingly recognised as a distinct and important property of CPM development, complementing conventional measures of model performance.

Several factors may influence of stability of model performance, including the size and heterogeneity of the development sample, the number of candidate predictors, predictor distributions, outcome prevalence, model complexity, and the modelling algorithm used [10–16]. Recent work has formalised the assessment of instability of individual predictions and proposed bootstrap-based procedures for evaluating it during model development [7, 9]. However, evidence remains limited on how specific modelling decisions affect the stability of individual predictions. One of the key modelling decisions in clinical prediction model development is how continuous predictors are included [6]. Continuous predictors are commonly modelled using approaches that either simplify, constrain, or flexibly estimate their association with the outcome [17]. These modelling choices differ in the extent to which they preserve information, allow non-linear relationships [17–19], and may affect prediction stability, particularly when the development sample size is limited.

This study examined how methods for modelling continuous predictors influence prediction stability in clinical prediction model development, using real clinical data as an empirical basis. Rather than identifying a single optimal method, it described patterns in prediction stability across different representations of continuous predictors and varying development sample sizes. By considering simple and more complex approaches, the study aimed to clarify how these choices may affect the reproducibility of individual risk estimates under certain development conditions.

## Methods

### Study design

This was an empirical methodological study designed to investigate the influence of continuous predictor modelling methods on the stability of individual predictions in clinical prediction model development. The study focused specifically on continuous predictors and evaluated whether different approaches for incorporating these predictors produced different levels of prediction stability across varying sample size scenarios. Prediction stability, assessed following bootstrap framework [9], was the primary focus of the study. Measures of discrimination and calibration were also examined.

### Data source

Rather than relying on simulated data, we used a real clinical dataset to provide empirical evidence under realistic development conditions. The dataset was obtained from a previously published clinical prediction model developed to predict hospital admission at triage in the emergency department of Lampang Hospital, Thailand, which included 19,418 emergency department patients [20]. The original model was developed using logistic regression and incorporated predictors from multiple domains.

### Candidate predictors

Six continuous predictors were pre-selected from the original dataset: age, mean arterial pressure, pulse rate, respiratory rate, oxygen saturation, and body temperature. None of the six candidate predictors had missing values; therefore, no imputation was performed.

### Scenario analyses by sample size and modelling approaches

Scenario analyses were conducted to examine how sample size and continuous predictor modelling methods influenced prediction stability. The base sample size was defined using the minimum sample size approach proposed by Riley et al [21]. For a binary prediction model with 12 parameters, assuming a C-statistic of 0.70 and an outcome prevalence of approximately 40%, the required sample size was 871 patients with 349 events, corresponding to approximately 29 events per parameter (EPP). The 12-parameter assumption, equivalent to a maximum of two parameters per predictor, ensured that the sample size was sufficient for modelling approaches requiring two degrees of freedom per continuous predictor, such as tertile categorisation or quadratic terms.

Five development sample size scenarios were then created by applying multiplicative factors of 0.5, 1, 2, 4, and 10 to the base sample size. For each scenario, a subsample was drawn from the original dataset using stratified random sampling. Within each sample size scenario, six methods for modelling continuous predictors were compared: dichotomisation at the median (DIC); tertile categorisation using cut-points at the 33rd and 67th percentiles (TER); linear terms (LIN); linear and quadratic terms (QUA); the multivariable fractional polynomial algorithm (MFP); and extreme gradient boosting (XGB) as a machine learning comparator. Logistic regression was used for the DIC, TER, LIN, QUA, and MFP approaches.

### Model evaluation

Prediction stability was the primary focus, with discrimination and calibration assessed as supporting measures. Discrimination was evaluated using the area under the receiver operating characteristic curve (AUC). Calibration was assessed using calibration-in-the-large (CITL) and calibration slope.

**Internal validation**

Internal validation was performed using bootstrap resampling with 200 replicates. In each iteration, a sample of the same size as the development dataset was drawn with replacement, and the full modelling process was repeated, including the specified method for handling continuous predictors. Each bootstrap model was then evaluated in both the bootstrap sample and the original development sample to estimate optimism in AUC, CITL, and calibration slope. The average optimism across all replicates was subtracted from the apparent estimates to obtain optimism-corrected performance measures.

**Prediction stability assessment**

Prediction stability was assessed using the same bootstrap resampling procedure as the internal validation. First, the model developed in the original development sample was used to estimate predicted probabilities for all individuals in that sample. In each bootstrap replicate, a new model was developed using the same modelling method and applied to the original development sample to generate bootstrap-based predicted probabilities.

For each individual, prediction instability was calculated as the absolute difference between the predicted probability from the bootstrap model and that from the original development model. These absolute differences were summarised using the mean absolute prediction error (MAPE). We also calculated the proportion of individuals with MAPE ≤5%, with a proportion of at least 90% considered to indicate acceptable prediction stability. In addition, optimism-corrected calibration slope and calibration-in-the-large (CITL) were used as supporting measures, with values between 0.95 and 1.05 for calibration slope and between −0.05 and 0.05 for CITL considered acceptable. These thresholds were pre-specified as pragmatic reference criteria rather than formal universal standards. They were informed by previous relevant works and by common interpretative principles in clinical prediction modelling [9, 22, 23].

### Software and computational resources

All analyses were conducted using R version 4.4.2 (R Foundation for Statistical Computing, Vienna, Austria) in RStudio (Posit Software, PBC, Boston, MA, USA). Calibration measures were estimated using the CalibrationCurves package. Multivariable fractional polynomial modelling was implemented using the mfp package, and extreme gradient boosting was implemented using the xgboost package. Figures were created using StataNow 19.5 (StataCorp, College Station, TX, USA). Bootstrap resampling was performed using the high-performance computing (HPC) resources of the Faculty of Medicine, Chiang Mai University, including the MedCMU-HPC-RAPTOR cluster.

## Results

The five development datasets included 437 to 8,739 patients, with 176 to 3,533 admission events and similar outcome prevalence across datasets (39.4%–42.0%) (Table 1). This corresponded to 29.3 to 588.8 EPP and 14.7 to 294.4 EPP under the maximum 12-parameter specification. Across bootstrap replicates, DIC and LIN consistently used six parameters, while QUA and TER used 12 parameters. The number of parameters selected by MFP varied across datasets, with the mean increasing from 8.0 at n = 437 to 11.9 at n = 8,739 (Supplementary Table S1).

Baseline characteristics were broadly comparable across sample size scenarios (Table 1), with similar marginal distributions of continuous predictors across the full dataset and subsampled datasets (Supplementary Figure S1). The smoothed predictor–outcome relationships suggested that the main patterns were generally preserved across datasets (Supplementary Figure S2).

**Table 1. Baseline characteristics across datasets.**

| | **0.5xBase** | **Base** | **2xBase** | **4xBase** | **10xBase** |
|---|---|---|---|---|---|
| Sample size | n=437 | n=874 | n=1748 | n=3496 | n=8739 |
| No. of events | 176 (40.3%) | 367 (42.0%) | 712 (40.7%) | 1378 (39.4%) | 3533 (40.4%) |
| Age (years) | 47.22 ± 21.87 | 50.28 ± 22.89 | 49.55 ± 22.00 | 49.61 ± 21.95 | 49.69 ± 22.14 |
| Mean arterial pressure (mmHg) | 99.40 ± 18.05 | 98.77 ± 17.05 | 97.61 ± 17.02 | 97.95 ± 17.30 | 97.89 ± 17.21 |
| Pulse rate (bpm) | 92.67 ± 19.26 | 91.02 ± 20.11 | 90.16 ± 18.78 | 90.21 ± 19.53 | 89.97 ± 18.93 |
| Respiratory rate (breaths/min) | 20.34 ± 3.30 | 20.46 ± 3.82 | 20.20 ± 3.20 | 20.20 ± 3.40 | 20.30 ± 3.44 |
| $SpO_2$ (%) | 97.49 ± 2.63 | 97.41 ± 3.42 | 97.39 ± 3.85 | 97.50 ± 3.02 | 97.43 ± 3.30 |
| Temperature (°C) | 36.82 ± 0.92 | 36.79 ± 0.91 | 36.73 ± 0.83 | 36.71 ± 0.80 | 36.73 ± 0.82 |

**Abbreviations:** $SpO_2$, peripheral oxygen saturation; bpm, beats per minute; SD, standard deviation. Values are presented as mean ± SD or n (%).

**Prediction stability across modelling scenarios**

Prediction stability increased with larger development sample sizes across all modelling methods (Figure 1-3). The proportion of individuals with MAPE ≤5% was lowest at n = 437 and exceeded the 90% stability criterion for all methods by n = 3,496. Median MAPE showed a corresponding decrease across sample size scenarios, with LIN having the lowest median MAPE throughout, decreasing from 0.041 at n = 437 to 0.009 at n = 8,739. XGB had the highest median MAPE in each scenario, decreasing from 0.077 to 0.019 (Supplementary Table S2).

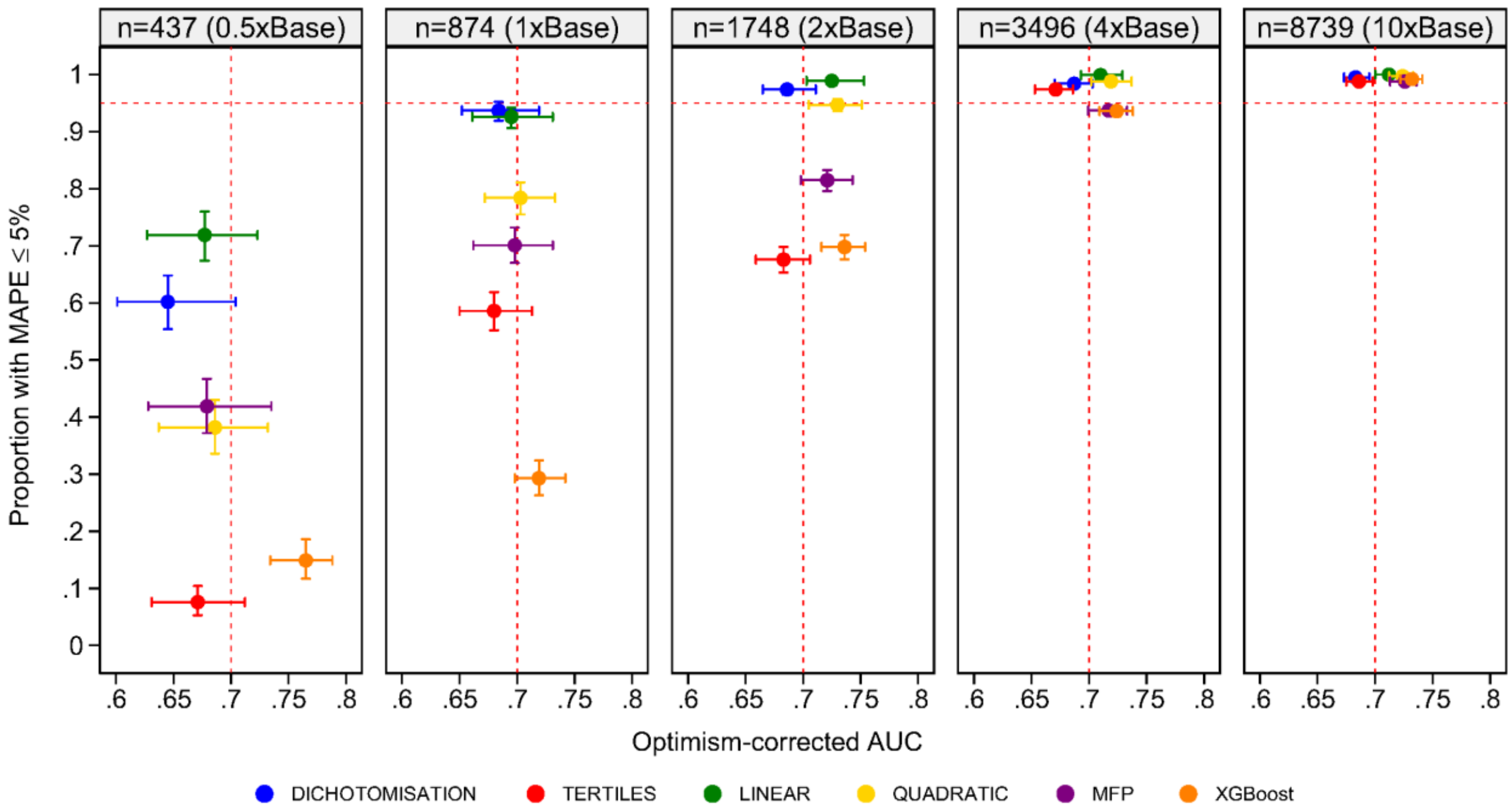

**Figure 1. Discriminative performance versus stability of six methods for modelling continuous predictors across different sample sizes.** Points represent the median optimism-corrected area under the receiver operating characteristic curve (AUC) and the proportion of models with mean absolute percentage error (MAPE) ≤5% across simulation runs. Horizontal and vertical error bars indicate the 2.5th–97.5th percentile intervals for AUC and the binomial 95% confidence interval for the proportion of bootstrap replication with MAPE ≤5%, respectively. Panels correspond to different sample size scenarios: n = 437 (0.5× base), n = 874 (1× base), n = 1748 (2× base), n = 3496 (4× base), and n = 8739 (10× base). Red dashed lines indicate reference thresholds of AUC = 0.70 and proportion with MAPE ≤5% = 0.90. **Abbreviations:** AUC; area under the receiver operating characteristic curve. MAPE; mean absolute percentage error. MFP; multivariable fractional polynomials.

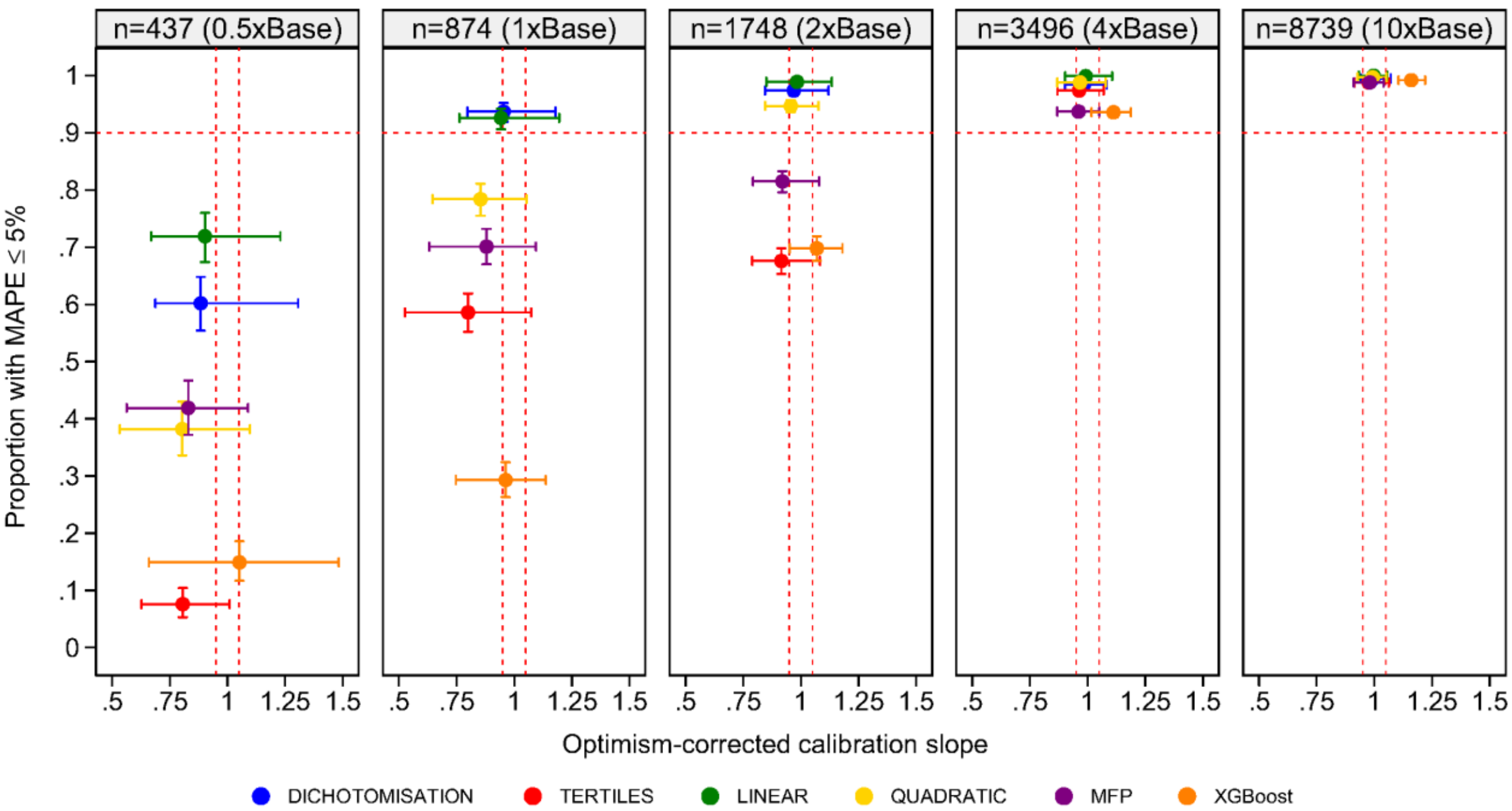

**Figure 2. Calibration slope versus stability for six methods for modelling continuous predictors across different sample sizes.** Points represent the median optimism-corrected calibration slope and the proportion of models with mean absolute percentage error (MAPE) ≤5% across simulation runs. Horizontal and vertical error bars indicate the 2.5th–97.5th percentile intervals for AUC and the binomial 95% confidence interval for the proportion of bootstrap replication with MAPE ≤5%, respectively. Panels correspond to different sample size scenarios: n = 437 (0.5× base), n = 874 (1× base), n = 1748 (2× base), n = 3496 (4× base), and n = 8739 (10× base). Red dashed lines indicate acceptable reference thresholds of calibration slope between 0.95 and 1.05 and proportion with MAPE ≤5% = 0.90. Abbreviations: MAPE; mean absolute percentage error. MFP; multivariable fractional polynomials.

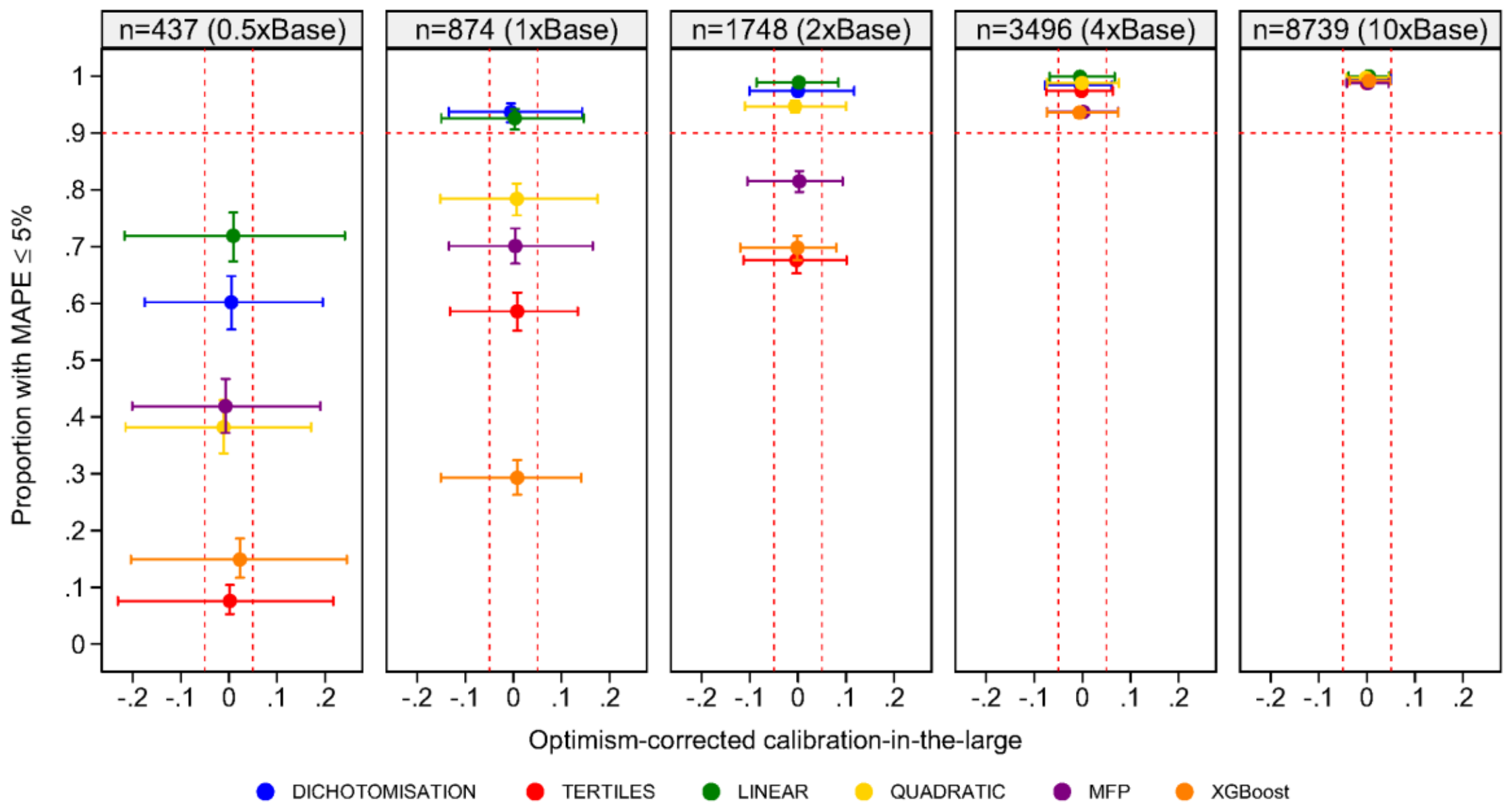

**Figure 3. Calibration-in-the-large versus stability for six methods for modelling continuous predictors across different sample sizes.** Points represent the median optimism-corrected calibration-in-the-large (CITL) and the proportion of models with mean absolute percentage error (MAPE) ≤5% across simulation runs. Horizontal and vertical error bars indicate the 2.5th–97.5th percentile intervals for AUC and the binomial 95% confidence interval for the proportion of bootstrap replication with MAPE ≤5%, respectively. Panels correspond to different sample size scenarios: n = 437 (0.5× base), n = 874 (1× base), n = 1748 (2× base), n = 3496 (4× base), and n = 8739 (10× base). Red dashed lines indicate acceptable reference thresholds of CITL between -0.05 and 0.05 and proportion with MAPE ≤5% = 0.90. **Abbreviations:** CITL; calibration-in-the-large. MAPE; mean absolute percentage error. MFP; multivariable fractional polynomials

Within each sample size scenario, stability differed by modelling method (Figure 1-3). At n = 437, no method achieved the pre-specified stability criterion. LIN showed the highest stability, with 71.9% of individuals having MAPE ≤5% (95% CI: 67.4%–76.0%), followed by DIC at 60.2% (95% CI: 55.4%–64.8%). The remaining methods ranged from 7.6% for TER to 41.9% for MFP. At n = 874, DIC and LIN achieved acceptable stability, with 93.7% and 92.6% of individuals having MAPE ≤5%, respectively. At n = 1,748, QUA also exceeded the stability criterion, reaching 94.7%, alongside DIC and LIN. By n = 3,496, all methods had achieved acceptable stability, with proportions ranging from 93.6% for XGB to 99.9% for LIN. At n = 8,739, all methods remained above the criterion, with proportions exceeding 98%.

**Discriminative performance**

Methods that retained continuous predictors as continuous terms generally had higher discrimination than categorisation approaches (Figure 1). XGB had the highest optimism-corrected AUC across all sample size scenarios, ranging from 0.719 at n = 874 to 0.765 at n = 437. Among the logistic regression approaches, QUA and MFP generally showed the highest AUCs when the sample size was at least 1,748, with values of approximately 0.72–0.73. DIC and TER had lower AUCs across most scenarios. At the largest sample size, the AUC was 0.683 for DIC and 0.686 for TER, compared with 0.712 for LIN, 0.724 for QUA, 0.726 for MFP, and 0.732 for XGB. Detailed optimism-corrected AUC values are presented in Supplementary Table S2.

**Model calibration**

For calibration slope, LIN was within or close to the acceptable range across most scenarios, with values from 0.943 at n = 874 to 0.997 at n = 8,739 (Figure 2). DIC and QUA also showed acceptable or near-acceptable calibration slopes once the sample size exceeded the smallest scenario. At n = 1,748, calibration slopes were 0.969 for DIC, 0.983 for LIN, and 0.956 for QUA. XGB showed calibration slopes above the acceptable range at larger sample sizes, increasing from 1.069 at n = 1,748 to 1.112 at n = 3,496 and 1.160 at n = 8,739. TER and MFP had calibration slopes below the acceptable range in smaller sample size scenarios, but were within the acceptable range at n = 3,496 and n = 8,739. Optimism-corrected CITL values were generally close to zero across all methods and sample size scenarios (Figure 3). Median CITL values ranged from −0.011 to 0.023 at n = 437 and from −0.002 to 0.005 at n = 8,739. Across all 30 scenarios, median CITL estimates were within the pre-specified acceptable range of −0.05 to 0.05. Detailed calibration results are shown in Supplementary Table S2.

**Discussion**

As expected, prediction stability improved as the development sample size increased. At the smallest sample size, none of the examined methods achieved the criterion, whereas all methods met this criterion by n = 3,496, which was the base reference size derived using the minimum sample size approach proposed by Riley et al [21]. This pattern is consistent with previous evidence that prediction models developed in smaller datasets are more vulnerable to instability, optimism, and variation in individual risk estimates. It also provides empirical support for the adequacy of this minimum sample size approach in achieving level-2 prediction stability in this setting [9]. However, stability still differed across the modelling methods within the same sample size scenario, indicating that sample size alone did not fully explain prediction stability.

Methods that retained continuous predictors on their original scale generally achieved a better balance between prediction stability and predictive performance than categorisation. Although dichotomisation (DIC) achieved acceptable stability from the base sample size onwards, it had lower discrimination than methods using continuous terms. Tertile categorisation (TER) performed less favourably, particularly in smaller datasets, despite having the same nominal number of parameters as quadratic modelling. These findings are consistent with previous concerns that categorising continuous predictors can reduce information, lower statistical efficiency, and introduce instability through cut-point selection [17, 18, 24]. Therefore, although dichotomisation appeared stable when the sample size was adequate, this should not be taken as evidence that continuous predictors should routinely be categorised.

Among the regression-based approaches, linear modelling (LIN) showed the most consistent stability. It achieved acceptable stability from the base sample size onwards and showed acceptable or near-acceptable calibration across most scenarios. Quadratic modelling (QUA) required a larger sample size before reaching the stability criterion, but provided higher discrimination than linear modelling once the sample size was sufficient. These results suggest that LIN may be preferable when sample size is limited, whereas QUA may offer a reasonable compromise between flexibility, discrimination, and stability in moderately larger datasets.

More flexible and complex approaches, such as MFP and XGB, required larger development samples before producing stable individual predictions. MFP did not meet the stability criterion until the base sample size, which may reflect the additional variability introduced by selecting functional forms and transformation powers. This is consistent with previous work showing that model-building procedures involving variable or function selection may be unstable in limited samples [12, 13]. In this study, such instability may also have been amplified by local fluctuations in the apparent predictor–outcome relationships in smaller subsamples. Although the subsamples broadly preserved the full-data distributions, the curves fluctuated more in smaller samples. Data-centric methods may be more sensitive to such variation because their functional forms, splits, or fitted structures can change according to the observed sample [25]. Although XGB showed the highest discrimination, it required larger samples to achieve stability and calibration. This suggests that strong discrimination does not necessarily translate into stable or well-calibrated individual risk estimates, particularly for more complex ML methods [26–28].

Prediction stability was not explained solely by the number of events per parameter (EPP). Although EPP is often used to assess sample size adequacy, it does not capture differences in model form, information preservation, or modelling complexity, all of which may influence prediction stability. This was reflected in the present study, where methods with the same nominal number of parameters, such as DIC and LIN or TER and QUA, showed different patterns of discrimination and stability. These findings support previous arguments that EPP rules are insufficient as a universal guide for sample size planning in prediction model development [21, 29, 30]. Instead, our findings support that the minimum sample size approach proposed by Riley et al. [21] may provide a more useful basis than EPP alone for assessing sample size adequacy in prediction model development.

This study has several strengths and limitations. A key strength is that it compared multiple approaches for modelling continuous predictors across increasing sample sizes within the same real clinical dataset. It also assessed stability at the level of individual predictions, rather than relying only on overall performance measures. By doing so, it provides practical evidence to help researchers balance predictive performance and stability when choosing how to model continuous predictors. However, there were several limitations. Firstly, this study was based on a single clinical dataset, which may limit generalisability. Secondly, we focused on simple models that included only continuous predictors and did not examine categorical predictors or interaction terms. More complex scenarios involving missing data, alternative outcome types, or regularisation methods, were beyond the scope of this study, although these methods may improve stability in some settings [16, 27]. Thirdly, dichotomisation and tertile categorisation used data-driven cut-points, which may not reflect settings where clinically defined thresholds are used. Finally, we did not evaluate other aspects of stability, such as calibration and classification instability. The assessment of prediction stability was limited to a bootstrap-based MAPE with pragmatic thresholds.

**Conclusions**

The present empirical analysis suggests that the way continuous predictors are modelled can influence the stability of individual predictions. LIN reached the stability criterion from the base sample size onwards, whereas QUA required a larger development sample and showed a reasonable balance between stability and discrimination. More flexible approaches, including MFP and XGB, required larger samples before achieving stable predictions, and XGB continued to show calibration concerns despite higher discrimination. For small or moderate development datasets, simpler modelling strategies, particularly LIN, may therefore be preferable when prediction stability is a priority. Because these findings are based on a single empirical dataset, further studies across different clinical settings, outcomes, and predictor structures are needed to confirm their generalisability.

**Acknowledgements**

This study was partially supported by Chiang Mai University and the Faculty of Medicine, Chiang Mai University. The authors also acknowledge the MedCMU-HPC-RAPTOR cluster, Faculty of Medicine, Chiang Mai University, for providing high-performance computing (HPC) resources for the bootstrap resampling analyses.

**Author contributions**

**Phichayut Phinyo**: Conceptualization, Methodology, Data curation, Formal analysis, Visualization, Investigation, Validation, Writing - Original Draft, Writing - Review & Editing, Software. **Pakpoom Wongyikul**: Conceptualization, Methodology, Validation, Writing - Review & Editing. **Noraworn Jirattikanwong**: Conceptualization, Methodology, Validation, Writing - Review & Editing. **Natthanaphop Isaradech**: Conceptualization, Methodology, Data curation, Formal analysis, Visualization, Investigation, Validation, Writing - Original Draft, Writing - Review & Editing, Software. **Wuttipat Kiratipaisarl**: Conceptualization, Methodology, Validation, Writing - Review & Editing. **Suppachai Lawanaskol**: Conceptualization, Methodology, Validation, Writing - Review & Editing. **Noppadon Seesuwan**: Conceptualization, Methodology, Validation, Writing - Review & Editing. **Wachiranun Sirikul**: Conceptualization, Methodology, Formal analysis, Visualization, Investigation, Validation, Writing - Original Draft, Writing - Review & Editing, Resources, Supervision, Project administration.

**Ethics approval**

This study was exempted from ethics review by the ethical committee of the Faculty of Medicine at Chiang Mai University (certificate of exemption no. 0267/2025) as it involved secondary analysis of de-identified data from a previously approved clinical prediction model development study.

**Funding**

This study receives no external funding.

**Conflict of interest**

None declared.

**Declaration of Generative AI and AI-assisted technologies in the writing process**

During the preparation of this manuscript, the authors used ChatGPT-5 and Gemini to assist with language editing, grammar correction, and clarity. All AI-assisted outputs were reviewed, edited, and verified by the authors, who take full responsibility for the content of the manuscript.

## Supplementary Appendix.

**Supplementary Table S1.** Number of model parameters across bootstrap models by sample size and modelling approach.

**Supplementary Table S2.** Performance and stability of six modelling approaches for continuous predictors across different sample size scenarios.

**Supplementary Figure S1.** Distribution of continuous predictors across the full dataset and five subsampled datasets.

**Supplementary Figure S2.** Predictor–outcome patterns for continuous predictors in the full dataset and development subsamples.

**Supplementary Table S1. Number of model parameters across bootstrap models by sample size and modelling approach**. Values summarise the number of model parameters across bootstrap-developed models within each sample size scenario and modelling approach. Parameter counts are not reported for XGB because model complexity is determined by tree structure and tuning parameters rather than regression coefficients.

| Sample size | Method | Number of models | Number of parameters | | | |
|---|---|---|---|---|---|---|
| | | | Mean | Minimum | Median | Maximum |
| 437 (0.5x) | DIC | 200 | 6.00 | 6 | 6 | 6 |
| | LIN | 200 | 6.00 | 6 | 6 | 6 |
| | QUA | 200 | 12.00 | 12 | 12 | 12 |
| | TER | 200 | 12.00 | 12 | 12 | 12 |
| | MFP | 200 | 7.97 | 7 | 8 | 10 |
| | XGB | 200 | — | — | — | — |
| 874 (1x) | DIC | 200 | 6.00 | 6 | 6 | 6 |
| | LIN | 200 | 6.00 | 6 | 6 | 6 |
| | QUA | 200 | 12.00 | 12 | 12 | 12 |
| | TER | 200 | 12.00 | 12 | 12 | 12 |
| | MFP | 200 | 8.39 | 7 | 8 | 12 |
| | XGB | 200 | — | — | — | — |
| 1748 (2x) | DIC | 200 | 6.00 | 6 | 6 | 6 |
| | LIN | 200 | 6.00 | 6 | 6 | 6 |
| | QUA | 202 | 12.00 | 12 | 12 | 12 |
| | TER | 200 | 12.00 | 12 | 12 | 12 |
| | MFP | 200 | 8.67 | 7 | 9 | 11 |
| | XGB | 200 | — | — | — | — |
| 3496 (4x) | DIC | 200 | 6.00 | 6 | 6 | 6 |
| | LIN | 200 | 6.00 | 6 | 6 | 6 |
| | QUA | 200 | 12.00 | 12 | 12 | 12 |
| | TER | 200 | 12.00 | 12 | 12 | 12 |
| | MFP | 200 | 9.91 | 7 | 10 | 12 |
| | XGB | 200 | — | — | — | — |
| 8739 (10x) | DIC | 200 | 6.00 | 6 | 6 | 6 |
| | LIN | 200 | 6.00 | 6 | 6 | 6 |
| | QUA | 200 | 12.00 | 12 | 12 | 12 |
| | TER | 200 | 12.00 | 12 | 12 | 12 |
| | MFP | 200 | 11.86 | 9 | 12 | 13 |
| | XGB | 200 | — | — | — | — |

**Abbreviations:** DIC, dichotomisation at the median; LIN, linear terms; QUA, quadratic terms; TER, tertile categorisation; MFP, multivariable fractional polynomials; XGB, extreme gradient boosting. XGB was not summarised by parameter count because model complexity is determined by tuning and tree structure rather than regression parameters.

**Supplementary Table S2. Performance and stability of six modelling approaches for continuous predictors across different sample size scenarios.** Values are presented as point (median) estimates with corresponding 2.5th–97.5th percentile intervals across bootstrap simulation replicates. For the proportion with MAPE ≤5%, 95% binomial confidence intervals are reported.

| Sample size | Method | Corrected AUC (2.5th–97.5th) | Corrected CITL (2.5th–97.5th) | Corrected Calibration slope (2.5th–97.5th) | MAPE (2.5th–97.5th) | Proportion with MAPE ≤5% (%) (95% CI) |
|---|---|---|---|---|---|---|
| 437 (0.5x) | DIC | 0.645 (0.601–0.704) | 0.005 (-0.175–0.195) | 0.884 (0.687–1.306) | 0.048 (0.038–0.073) | 0.602 (0.554–0.648) |
| | TER | 0.671 (0.631–0.712) | 0.002 (-0.231–0.217) | 0.806 (0.627–1.009) | 0.070 (0.043–0.103) | 0.076 (0.053–0.104) |
| | LIN | 0.677 (0.627–0.723) | 0.009 (-0.217–0.241) | 0.902 (0.669–1.229) | 0.041 (0.026–0.082) | 0.719 (0.674–0.760) |
| | QUA | 0.686 (0.637–0.732) | -0.011 (-0.215–0.171) | 0.804 (0.533–1.097) | 0.055 (0.029–0.117) | 0.382 (0.336–0.430) |
| | MFP | 0.679 (0.628–0.735) | -0.007 (-0.201–0.190) | 0.830 (0.563–1.089) | 0.054 (0.030–0.137) | 0.419 (0.372–0.467) |
| | XGB | 0.765 (0.734–0.788) | 0.023 (-0.204–0.245) | 1.052 (0.659–1.482) | 0.077 (0.038–0.117) | 0.149 (0.117–0.186) |
| 874 (1x) | DIC | 0.684 (0.652–0.719) | -0.005 (-0.134–0.143) | 0.954 (0.797–1.179) | 0.034 (0.025–0.072) | 0.937 (0.919–0.952) |
| | TER | 0.680 (0.650–0.713) | 0.008 (-0.132–0.134) | 0.800 (0.526–1.073) | 0.048 (0.029–0.119) | 0.586 (0.552–0.619) |
| | LIN | 0.695 (0.661–0.731) | 0.003 (-0.150–0.146) | 0.943 (0.762–1.195) | 0.028 (0.018–0.060) | 0.926 (0.906–0.942) |
| | QUA | 0.703 (0.672–0.733) | 0.007 (-0.152–0.175) | 0.854 (0.646–1.053) | 0.037 (0.020–0.098) | 0.784 (0.755–0.811) |
| | MFP | 0.698 (0.662–0.731) | 0.004 (-0.134–0.165) | 0.880 (0.632–1.094) | 0.041 (0.025–0.110) | 0.701 (0.670–0.732) |
| | XGB | 0.719 (0.698–0.742) | 0.008 (-0.151–0.141) | 0.963 (0.747–1.137) | 0.059 (0.029–0.092) | 0.293 (0.263–0.324) |
| 1748 (2x) | DIC | 0.686 (0.665–0.711) | 0.000 (-0.101–0.117) | 0.969 (0.845–1.120) | 0.024 (0.017–0.068) | 0.974 (0.965–0.981) |
| | TER | 0.683 (0.659–0.706) | -0.003 (-0.113–0.101) | 0.915 (0.789–1.082) | 0.041 (0.024–0.091) | 0.676 (0.653–0.698) |
| | LIN | 0.725 (0.703–0.753) | 0.002 (-0.086–0.084) | 0.983 (0.850–1.134) | 0.020 (0.012–0.042) | 0.989 (0.983–0.993) |
| | QUA | 0.730 (0.705–0.751) | -0.006 (-0.110–0.100) | 0.956 (0.845–1.076) | 0.024 (0.013–0.060) | 0.947 (0.936–0.957) |
| | MFP | 0.721 (0.698–0.743) | 0.003 (-0.105–0.093) | 0.920 (0.792–1.079) | 0.034 (0.019–0.101) | 0.815 (0.796–0.833) |
| | XGB | 0.736 (0.716–0.754) | -0.001 (-0.120–0.080) | 1.069 (0.951–1.180) | 0.042 (0.020–0.072) | 0.698 (0.676–0.719) |
| 3496 (4x) | DIC | 0.687 (0.670–0.703) | -0.003 (-0.078–0.061) | 0.984 (0.902–1.083) | 0.016 (0.012–0.037) | 0.984 (0.979–0.988) |
| | TER | 0.671 (0.653–0.686) | -0.002 (-0.075–0.063) | 0.963 (0.869–1.069) | 0.025 (0.015–0.051) | 0.974 (0.968–0.979) |
| | LIN | 0.710 (0.693–0.729) | -0.005 (-0.068–0.067) | 0.992 (0.903–1.106) | 0.013 (0.008–0.029) | 0.999 (0.997–1.000) |
| | QUA | 0.719 (0.702–0.737) | -0.001 (-0.074–0.076) | 0.967 (0.868–1.078) | 0.018 (0.009–0.043) | 0.988 (0.984–0.992) |

| Sample size | Method | Corrected AUC (2.5th–97.5th) | Corrected CITL (2.5th–97.5th) | Corrected Calibration slope (2.5th–97.5th) | MAPE (2.5th–97.5th) | Proportion with MAPE ≤5% (%) (95% CI) |
|---|---|---|---|---|---|---|
| | MFP | 0.717 (0.699–0.733) | 0.002 (-0.074–0.074) | 0.961 (0.868–1.051) | 0.023 (0.011–0.069) | 0.937 (0.929–0.945) |
| | XGB | 0.724 (0.709–0.738) | -0.006 (-0.074–0.074) | 1.112 (1.017–1.188) | 0.029 (0.016–0.058) | 0.936 (0.927–0.944) |
| 8739 (10x) | DIC | 0.683 (0.673–0.695) | -0.002 (-0.043–0.050) | 0.996 (0.941–1.071) | 0.010 (0.007–0.027) | 0.995 (0.994–0.997) |
| | TER | 0.686 (0.675–0.698) | 0.000 (-0.041–0.045) | 0.975 (0.914–1.063) | 0.015 (0.009–0.028) | 0.988 (0.985–0.990) |
| | LIN | 0.712 (0.700–0.722) | 0.004 (-0.038–0.046) | 0.997 (0.933–1.057) | 0.009 (0.005–0.019) | 1.000 (0.999–1.000) |
| | QUA | 0.724 (0.712–0.733) | -0.002 (-0.043–0.045) | 0.993 (0.928–1.053) | 0.011 (0.006–0.027) | 0.997 (0.996–0.998) |
| | MFP | 0.726 (0.713–0.736) | 0.002 (-0.042–0.045) | 0.980 (0.911–1.040) | 0.014 (0.008–0.039) | 0.988 (0.986–0.990) |
| | XGB | 0.732 (0.722–0.741) | 0.005 (-0.036–0.049) | 1.160 (1.105–1.220) | 0.019 (0.010–0.042) | 0.992 (0.989–0.993) |

**Abbreviations:** AUC; area under the receiver operating characteristic curve, CITL; calibration-in-the-large, MAPE; mean absolute percentage error, CI; confidence interval, DIC; dichotomisation of the continuous predictor, LIN; linear effect of the predictor, MFP; multivariable fractional polynomial modelling, QUA; quadratic transformation of the predictor, TER; categorisation into tertiles, XGB; extreme gradient boosting.

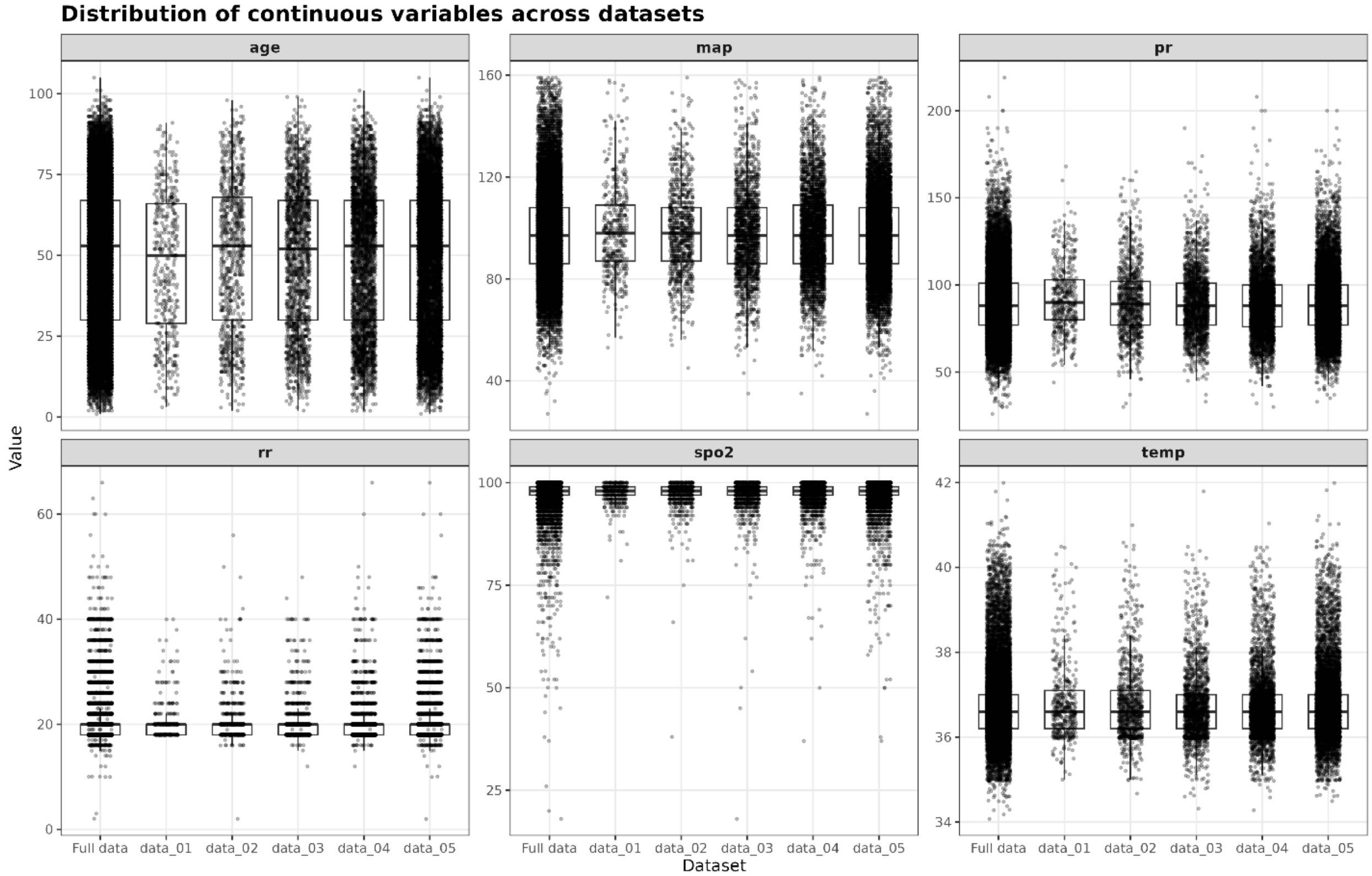

**Supplementary Figure S1. Distribution of continuous predictors across the full dataset and five subsampled datasets.** Boxplots with overlaid jittered points show the distributions of six continuous predictors in the full dataset and five subsampled datasets of varying sizes. The centre line of each box represents the median, the lower and upper bounds of the box represent the interquartile range, and whiskers represent the range within 1.5 times the interquartile range. Individual observations are shown as semi-transparent points. The figure was used to visually assess whether the marginal distributions of continuous predictors were broadly preserved across subsampled datasets. **Abbreviations:** MAP, mean arterial pressure; PR, pulse rate; RR, respiratory rate; $SpO_2$, peripheral oxygen saturation; Temp, body temperature.

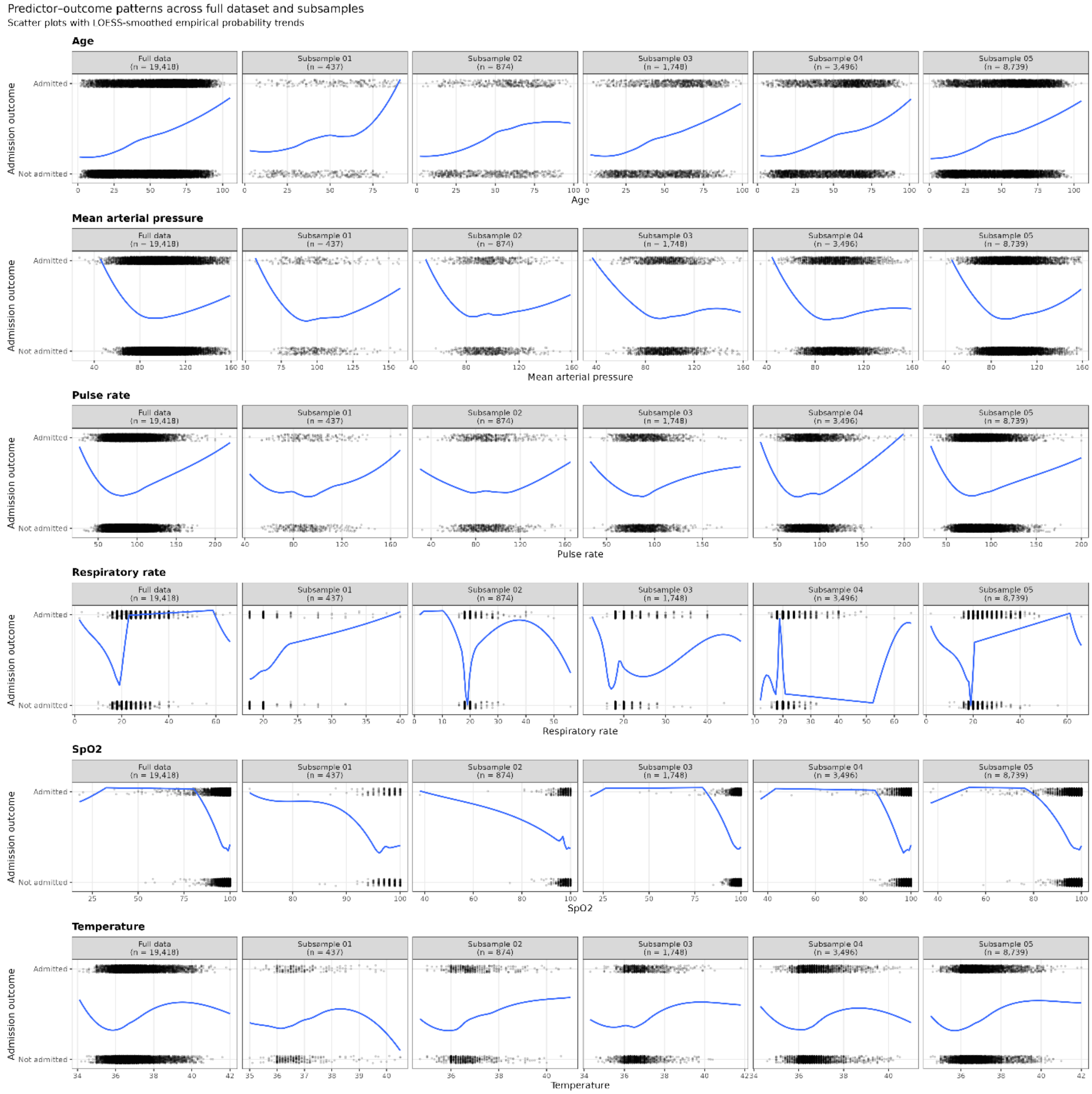

**Supplementary Figure S2. Predictor–outcome patterns for continuous predictors in the full dataset and development subsamples.** Scatter plots showing the empirical relationship between each continuous predictor and hospital admission in the full dataset and across five development subsamples. Each row represents one predictor, and each column represents the full dataset or one subsample size scenario. Grey points indicate individual patient observations, with admission outcome coded as not admitted or admitted. Blue lines represent locally estimated scatterplot smoothing (LOESS) curves showing the empirical probability trend of admission across the observed range of each predictor. The figure illustrates that predictor–outcome patterns varied across subsamples, particularly in smaller datasets, with smoother and more consistent patterns observed as sample size increased. **Abbreviations:** LOESS, locally estimated scatterplot smoothing; $SpO_2$, peripheral oxygen saturation.